\documentstyle[11pt,newpasp,twoside,psfig]{article}
\index{instructions}
\index{guidelines}

\def\edcomment#1{\iffalse\marginpar{\raggedright\sl#1\/}\else\relax\fi}
\reversemarginpar

\begin{document}
\title{The $\sigma$ problem of the Crab pulsar wind}
 \author{J.G. Kirk}
\affil{Max-Planck-Institut f\"ur Kernphysik, Postfach 10 39 80, 
69029~Heidelberg, Germany}
\author{O. Skj{\ae}raasen}
\affil{Observatoire de Strasbourg, 11 rue de l'Universit\'e, F--67000
 Strasbourg, France}

\begin{abstract}
The conversion of the Crab pulsar wind from one dominated by Poynting
flux close to the star to one dominated by particle-born energy at the
termination shock is considered. The idea put forward by
Coroniti (1990) and criticised by Lyubarsky \& Kirk (2001) 
that reconnection in a striped wind is responsible, is generalised 
to include faster prescriptions for the a~priori unknown dissipation
rate. Strong acceleration of the wind is confirmed, and the higher
dissipation rates imply complete conversion of Poynting flux into
particle-born flux within the unshocked wind.
\end{abstract}

\section{Introduction}
A puzzling feature of the relativistic MHD wind of the Crab pulsar is
that it appears to arrive at its termination shock as a particle
dominated flow (Rees \& Gunn 1974; Kennel \& Coroniti 1984; Emmering
\& Chevalier 1987). This is difficult to understand, because the wind almost
certainly carries its energy by Poynting flux close to the star, and,
given that
relativistic MHD winds collimate only weakly (Beskin, Kuznetsova \&
Rafikov 1998;
Chiueh, Li, \& Begelman 1998; Bogovalov \& Tsinganos 1999; Lyubarsky
\& Eichler 2001, Lyubarsky 2002)
it is not obvious how the transformation 
to particle-born flux
can occur, or why it is
apparently not observed. The acceleration intrinsic to axisymmetric 
{\em force-free} winds (Buckley 1977; Contopoulos \& Kazanas 2002)
does not alleviate the difficulty, since the flow is still
magnetically dominated outside the sonic point where such solutions lose
their validity. 

An obliquely rotating neutron star drives a wind which, near the
equatorial plane, consists of alternating stripes of magnetic field of
opposite polarity. The magnetic field is linked alternately 
to the open field lines emerging from different magnetic hemispheres,
as these are swept around the rotational equator. This led Coroniti
(1990) to suggest that reconnection or, more accurately, dissipation
at the current sheet separating the alternating polarities, might
effect the required conversion. He estimated the dissipation rate by
assuming the thickness of the current sheet was approximately equal to
the gyro radius of the hot particles it contained and found that the
magnetic field could be annihilated before the wind reached the
termination shock. However, Coroniti's treatment neglected the
acceleration of the wind flow. Using the same micro-physics estimate 
for dissipation, Lyubarsky \& Kirk (2001) showed that this effect 
results in an annihilation rate far too small to be effective.    

Kirk \& Skj{\ae}raasen (2003) recently identified 
the dissipation estimate of Coroniti as a lower limit and 
examined the effect of alternative micro-physics prescriptions,
including an upper limit on the dissipation rate. Their findings,
which we paraphrase in this paper, re-open the possibility that the
conversion of Poynting flux to particle-born flux occurs through
dissipation of magnetic field energy at the current sheet of a striped
wind.

\section{Small wavelength approximation and dissipation prescriptions}
The approach of Lyubarsky \& Kirk (2001) and Kirk \& Skj{\ae}raasen
(2003) is based on computing the slow evolution of an entropy wave as it
is convected outwards by the stellar wind. The method employs a
separation of length scales into a short one, identified with 
the wavelength of the magnetic field reversals, and a long one,
identified with the radius, i.e., the scale on which the 
density changes as a result of the spherically expanding flow. 
The equations of conservation of particles and energy/momentum 
are then ordered in the small
parameter formed from the ratio of these lengths. The entropy wave
whose evolution is sought can be described in several ways. One is in
terms of the parameters which describe the {\em Harris sheath}
(Harris 1962; Hoh 1968); another is in terms of the parameters of 
two uniform fluid layers --- one hot
and unmagnetized, the other cold and magnetized --- separated from
each other by discontinuities in density, velocity and magnetic field,
but having equal pressures. Each of these 
configurations satisfies the zeroth order 
conservation equations, indicating that the wave does not evolve on
this short length-scale.

On the long length scale, however, evolution of the wave is possible. It is
computed by regularising the expansion, i.e., by demanding that the secular
terms in the first-order equations vanish. The conservation equations
by themselves are not sufficient to solve for all the parameters of
the wave as a function of radius --- an additional constraint is
required, reflecting the need for new physics input to determine the
rate at which dissipation proceeds. In MHD simulations, the new
input needed to describe dissipation is a functional relation between
the current and the electric field: an {\em anomalous
  conductivity}. For our computations, however, it suffices to
prescribe one of the wave parameters (the thickness of the Harris sheath,
for example) in terms of the others. 

The Coroniti (1990) prescription 
implies a relative drift speed of the oppositely charged species
roughly equal to that of light. In other words, there exists no Harris
equilibrium for a sheath thinner than that assumed by Coroniti. The 
thinner the sheet, the slower the dissipation, so that Coroniti's
prescription provides a lower limit on the dissipation rate. Kirk \&
Skj{\ae}raasen (2003) consider two alternative prescriptions: that the sheath
thickness is governed by the (linear) growth rate of the tearing
instability, as suggested by Lyubarsky (1996), and that the expansion
speed of the sheath is limited to its internal sound speed. The
motivation for the latter choice stems from the observation that only
the oscillating component of the magnetic field in the entropy wave
can be dissipated. This suggests that the boundaries of the hot fluid
must be in causal contact in order 
to guarantee that they advance into regions of 
opposite magnetic polarity at equal speeds. Thus, 
expansion at the internal sound speed
provides an upper limit to the dissipation rate.  

\begin{figure}
\hfil\psfig{file=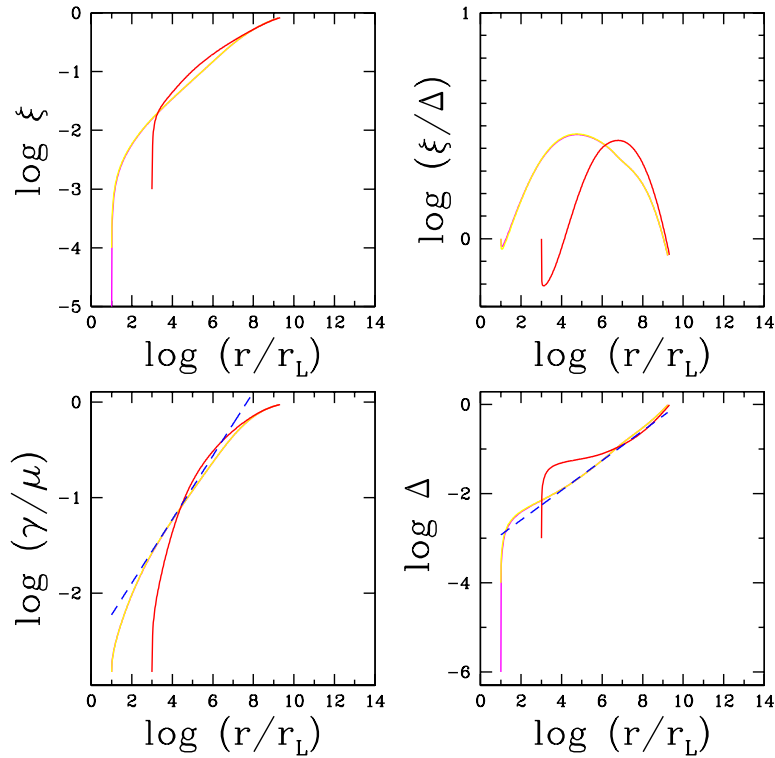,width=0.8\textwidth}\hfill
\caption{The Crab pulsar wind with maximum rate of dissipation (sonic
  sheath expansion) and particle
  injection rate into the Nebula of
  $3\times10^{40}\,\textrm{s}^{-1}$. Shown 
as functions of $r/r_{\rm L}$, (where $r_{\rm L}$ is the radius of the
  light cylinder)
are the ratio $\xi$ of the particle 
  density in the current sheet to that in the magnetized wind, the
  Lorentz factor $\gamma$ in units of its maximum value $\mu$
  ($=2\times10^4$
in this case), the thickness $\Delta$ of the sheath in units of $\pi
  r_{\rm L}$ and the ratio $\xi/\Delta$. The termination shock is located
  at $\log(r/r_{\rm L})\approx 9$. 
}
\end{figure}

\section{Implications for the Crab pulsar wind}
The dissipation of the magnetic field in the wind of the Crab pulsar
depends on the particle load carried by the wind. For a particle loss
rate of $\dot{N}<6\times10^{38}\,\textrm{s}^{-1}$, the theoretical
upper limit on the dissipation rate lies below the lower limit
given by the Coroniti prescription, so that the
treatment is inconsistent. In this case, slow evolution of the entropy
wave is impossible. Presumably the system adjusts by either becoming
unsteady or by steadily converting the flux into a mode which
propagates in the fluid frame (e.g., an electromagnetic mode).
For 
$6\times10^{38}\,\textrm{s}^{-1}<\dot{N}<3\times10^{40}\,\textrm{s}^{-1}$
none of the three dissipation prescriptions succeeds in converting
significant amounts of Poynting flux before the termination shock is
encountered. For $\dot{N}>3\times10^{40}\,\textrm{s}^{-1}$, however,
dissipation at the fastest permitted rate, as depicted in Fig.~1,
results in complete conversion before the termination shock is
encountered. 
This
rather high level of pair injection into the Nebula cannot be
explained in terms of current theories of cascade formation in the
inner magnetosphere (Hibschmann \& Arons 2001), but is nevertheless
consistent with the observed number of radio-emitting electrons    
(Gallant et al 2002).

Can the conversion of Poynting flux to
particle-born flux be observed directly? 
Arons (1979) noticed that conditions in the wind will
make such a signal pulse at the rotation
frequency of the star. The obvious implication that the optical to
gamma-ray pulses are indeed the signature of this conversion is
further supported by the approximate 
correspondence of the sheet geometry to the
observed pulse form (Kirk, Skj{\ae}raasen, \& Gallant 2002).

\acknowledgments
We thank Yves Gallant and Yuri Lyubarsky for helpful
discussions.

\end{document}